\documentstyle[11pt,aaspp4]{article} 
\newcommand{\kms}{km~s$^{-1}$}

\newcommand{\subsun}{\mbox{$_{\odot}$}}
\newcommand{\etal}{{\it et al.\/}}
\newcommand{\teff}{$T_{\mbox{\scriptsize eff}}$}
\newcommand{\grav}{log($g$)}

\lefthead{Cohen {\it et al.} }
\righthead{Extremely Metal Poor Stars}

\begin{document}

\title{Stellar Archaeology: a Keck Pilot Program on Extremely Metal-Poor 
Stars From the Hamburg/ESO Survey. I Stellar Parameters\altaffilmark{1}}

\author{Judith G. Cohen\altaffilmark{2}, Norbert Christlieb\altaffilmark{3}, 
Timothy C. Beers\altaffilmark{4}, Raffaele Gratton\altaffilmark{5} \& 
Eugenio Carretta\altaffilmark{5} }

\altaffiltext{1}{Based in large part on observations obtained at the
	W.M. Keck Observatory, which is operated jointly by the California 
	Institute of Technology, the University of California
        and NASA,}
\altaffiltext{2}{Palomar Observatory, Mail Stop 105-24,
	California Institute of Technology, Pasadena, CA \, 91125,
        jlc@astro.caltech.edu}
\altaffiltext{3}{Hamburger Sternwarte, Gojenbergsweg 112, D-21029 
Hamburg, Germany, nchristlieb@hs.uni-hamburg.de}
\altaffiltext{4}{Department of Physics and Astronomy, Michigan State University,
East Lansing, Michigan 48824-1116, beers@pa.msu.edu}
\altaffiltext{5}{INAF - Osservatorio Astronomico di Padova, Vicolo dell'Osservatorio 5,
        35122, Padova, Italy, carretta, gratton@pd.astro.it}

\begin{abstract}
In this series of two papers we present a high
dispersion spectroscopic analysis of 8 candidate extremely metal poor
stars selected from the Hamburg/ESO Survey and of 6 additional
very metal poor stars.  We demonstrate
that with suitable vetting using moderate-resolution spectra the yield of 
this survey for stars with
[Fe/H] $\le -3.0$ dex is very high; 
three out of the eight 
stars observed thus far at high resolution from the HES are actually that metal poor, 
three more have [Fe/H] $\le -2.8$ dex, and
the remainder are
only slightly more metal rich.  In preparation for a large scale
effort to mine the Hamburg/ESO Survey database for such stars 
about to get underway, we lay out in this paper the basic principles
we intend to use to determine in a uniform way the stellar parameters 
\teff, \grav, and reddening.

\end{abstract}


\section{Introduction}

The most metal deficient stars in the Galaxy provide crucial evidence
on the early epoch of the formation of our Galaxy, the chemical evolution of
the Galaxy, the environment in which various elements
were produced, the production of elements in the Big Bang, 
the age of the Galaxy, the relationship
between the halo field stars and the galactic globular clusters, etc.
We consider only stars with [Fe/H] $\le -3$ dex 
($<$1/1000 of the solar metallicity) to be extremely metal poor
(henceforth EMP)
and most useful for discussion of such issues.  These EMP stars provide
us with an opportunity to study in detail the local equivalent of the
high redshift Universe.

The major existing survey for very metal poor stars is the HK survey
(previously referred to as the Preston-Shectman survey) described in detail by
Beers, Preston \& Shectman (1985, 1992).  This survey produced 
a list of $\sim$10,000 candidates for metal poor stars.
Extensive follow up studies,
which continue to this date, now include moderate-resolution
($\Delta(\lambda) \sim 1 - 2$\AA) spectroscopy of about 5000 stars
and broad- and narrow-band photometry for some 3000 stars.
From this work one can cull a sample of roughly 1000 stars
with [Fe/H] $\le -2.0$, but, as summarized by Beers (1999), only roughly 100
are known to be extremely metal poor.  High-dispersion analyses exist for only
a few stars with [Fe/H] $\le -3.5$, and the signal to noise ratio of these
spectra are in general modest (${\sim}30$ to 40), a situation which only very
recently has been partially remedied for five giants through the work of
Norris, Ryan \& Beers (2001), and for the most metal-poor dwarf presently
known, CS~22876-032 (Norris, Beers \& Ryan 2000).
 
McWilliam (1997) reviews the key results from abundance
analyses for the most metal poor stars known from the HK survey.
The elements Al, Sr, Ba, Cr, Mn, Co  show a sudden change in the 
slope of [X/Fe] versus [Fe/H] near
[Fe/H] = $-2.4$ dex.  Other elements, including Ba and Sr,
appear to show considerable scatter from star-to-star
at very low [Fe/H], perhaps indicating that 
the yield from individual or at most a few supernova events produced
the metals seen in these EMP stars.
However, the total sample of stars
involved is 33 in the McWilliam \etal\ (1995) study and 19 in Ryan \etal\ (1996),
with some overlap between these two samples, and many of these do not
fit the strict definition of EMP stars adopted here.

The large scatter in the element-to-element abundance ratios observed 
amongst EMP stars might reveal a wealth of data for nucleosynthesis. However, 
a proper understanding of this scatter requires large samples. 
Furthermore, interesting new types of stars such as those extremely rich in 
r-process elements (useful for nucleocosmochronology) or extreme CH-stars 
(used to to study the s-process) have been found based on the HK survey,
but the number known in either case is very small, and urgently needs
augmentation.

      The Hamburg/ESO survey (HES) is an objective prism survey
      primarily targeting bright quasars (Wisotzki \etal\ 1996, 2000).
However, because its
      spectral resolution is typically 15\,{\AA} FWHM at H$\gamma$, it
      is also possible to efficiently select a variety of interesting
      \emph{stellar} objects in the HES (Christlieb 2000; Christlieb et
      al. 2001a,b), among them EMP stars.
The HES is based on automated scans of
   objective prism plates. With a nominal area of 9,575 deg$^2$, it
   covers the entire southern extragalactic sky ($|b| \ge 30$ deg,
and $\delta <+2.5$ deg).
   The HES limiting magnitude for metal-poor stars is $B\sim 17.5$.  The
survey is now completed and the HES database contains digitized
objective prism spectra for about 4 million stars.

A comparison of the EMP stars likely to be found with
this survey as compared with the HK survey is given
in Christlieb \& Beers (2000).
The HES has a number of crucial advantages
over the HK survey which should lead to a major increase
in the samples of EMP stars and hence, with suitable
follow up observations, in our knowledge of their properties.
The advantages of the HES include a deeper magnitude limit, broader spectral
coverage, and an automated selection
from digitized scans of the plates.  The latter
ensures a selection of EMP 
stars which can be independent of stellar effective temperature.  

The existence of a new
list of candidates for EMP stars 
with [Fe/H] $< -3$ dex 
selected in an automated and unbiased manner
from the HES, coupled with
the very large collection area and efficient high resolution Ech\`elle
spectrographs of 8--10 m telescopes such as Keck+HIRES or VLT+UVES,
offers the possibility for a very large
increase in the number of EMP stars known
and in our understanding of their properties.  
The present pair of papers (this paper and Paper II, Carretta \etal\ 2001)
describes 
our very successful pilot project with HIRES
at the Keck Observatory in the fall of 2000 
to determine the effective yield of the HES for
EMP stars through high dispersion abundance analyses of a sample of
stars selected from the HES.
Our work complements and extends that
of Depagne, Hill, Christlieb \& Primas (2000), who analyzed
two stars from the HES as part of the UVES science verification at the VLT.

This success has provided the stimulus for the creation of a long term
large scale international effort to mine the HES for EMP stars
(the 0Z Project), which is
now getting underway, and which will complement the efforts of the
ESO-VLT large program led by Cayrel (see, e.g. Cayrel \etal\ 2001) dedicated to
studying the HK survey stars themselves in more detail.  In this
pair of papers we strive to lay down at least some of the procedures
that we will follow in our long term effort, whose goal is to 
dramatically increase the sample of known and well studied EMP stars.  
This first paper dedicated to determination of the stellar parameters.

\section{The Sample of Stars Observed \label{sample} }

      Selection of EMP stars in the HES is carried by automatic spectral
      classification, using classical statistical methods (Christlieb 2000).
      As described in Christlieb \etal\ (2001a), $B-V$ colors can be estimated
      directly from the digital HES spectra with an accuracy of $\sim
      0.1$\,mag, so that these samples can be selected not only on the basis of
      spectroscopic criteria but also with restrictions on $B-V$ color.

      The principal spectroscopic criterion used for sample selection for EMP
      stars is the same as that used by the HK project, the absence/weakness
      of the 3933\,{\AA} line of Ca~II.
A visual check of the HES spectrum is then made
to eliminate the small fraction of spurious objects (plate defects,
misidentifications, etc.)
that pass the automatic selection criteria.  
The details of this procedure will be
discussed elsewhere (Christlieb \etal\, in preparation).

The results from follow-up observations of
      several hundred metal-poor candidates from the HES will be described
      elsewhere (Christlieb \etal\, in preparation). Here we only note that
      in the HES a very high selection efficiency 
for metal-poor stars has been achieved: $\sim 60$\,\% of the candidate 
EMP turnoff stars observed at moderate-resolution were
      confirmed to have $\mbox{[Fe/H]}<-2.0$, compared to $\sim 30$\,\% in
      the HK survey (Beers 2000). For the cooler giants, having a stronger
      Ca~K line at given metallicity than turnoff stars, we expect the
      selection efficiency to be even higher due to the still detectable
      Ca~K line at very low metallicities.  This is currently
under investigation.

The present
      sample was selected from the HES database 
to have $0.3<B-V<0.5$ to focus on main sequence
      turnoff stars.
      The pool of candidates in the turnoff region consists of those stars
      whose line strengths are below the limit of detection for the HES.  
This limit corresponds to a
      metallicity which exceeds the upper limit 
adopted here for EMP stars, and hence the
      initial samples are dominated by stars that are very metal poor, but
      more metal rich than those we seek. Thus, to make the best use of the
      limited observing time available on the largest telescopes, these
      candidates from the HES database must first be verified through 
moderate-resolution ($\sim$1--2\,{\AA}) follow-up spectroscopy at 4-m class
      telescopes.  
This procedure selects out the genuine EMP stars from the much more numerous
stars of slightly higher metallicity -- of interest in their own right -- but
not relevant for our present study.  It
is the overall efficiency of this multi-stage selection 
process for isolating genuine EMP stars which we seek to establish with our
pilot project.

\section{HIRES Observations \label{hires} }

Once the list of vetted candidates for EMP candidates from the HES
database has been created, observations
at high dispersion for a full scale abundance analysis can be
undertaken.

The sample of stars studied here includes seven previously vetted
EMP
main-sequence turnoff candidates and one giant candidate from the HES.
Their coordinates are given in Table~\ref{table_coords}. One
of these turned out to be a re-discovery of a star from the HK Survey
(HE~2344--2800 = CS 22966--048).  For comparison
and calibration, three bright,
well-studied very metal-poor stars were observed, 
two of which (G139-8 and BD+3~740) are known to be 
Li-deficient from the work of Norris, Ryan \& Beers (2000).
Three candidate EMP stars from the HK survey were also included.
Observations were carried out with the HIRES spectrograph at Keck I
on two nights in September 2000.
A spectral resolution of 45,000
was achieved using a 0.86 arcsec wide slit projecting to 3 pixels in
the HIRES focal plane CCD detector.
The spectra cover the region from 3870 to 5400\,{\AA} with essentially
no gaps.  Each exposure for the HES stars was broken up into
1200 sec segments.  
The spectra were exposed until a SNR of 100 per spectral
resolution element in the continuum at
4500\,{\AA} was achieved.  This SNR calculation utilizes only
Poisson statistics, ignoring issues of cosmic ray removal,
night sky subtraction, flattening, etc.   The observations
were carried out with the instrument rotator fixed in the
vertical position with respect to the horizon (i.e. at the parallactic angle).
As suitable candidates are quite far apart on the sky
compared to the 11 arcsec long slit, only
a single object can be observed at once.

The exposure times and signal to noise ratios per spectral
resolution element in the continuum are
given in Table~\ref{table_phot}.  (The gain setting of the HIRES
CCD detector is 2.4 e$^-$/ADU; to reach the desired SNR of 100 
per spectral resolution element requires 
1390 ADU/pixel.)
These spectra were taken
during a period when the Keck I mirror segments had not been
aluminized for more than two years, and the overall
throughput at these blue wavelengths needed improvement.

The spectra from both nights were reduced 
using the suite of routines for analyzing Ech\`elle
spectra written by McCarthy (1988) within the Figaro image processing package
(Shortridge 1988).  The stellar data are flat fielded with quartz lamp
spectra, thereby removing most of the blaze profile, and the results are
normalized to unity by fitting a 6th-order polynomial to line-free regions of
the spectrum in each order.

\begin{figure}
\epsscale{1.0}
\plotone{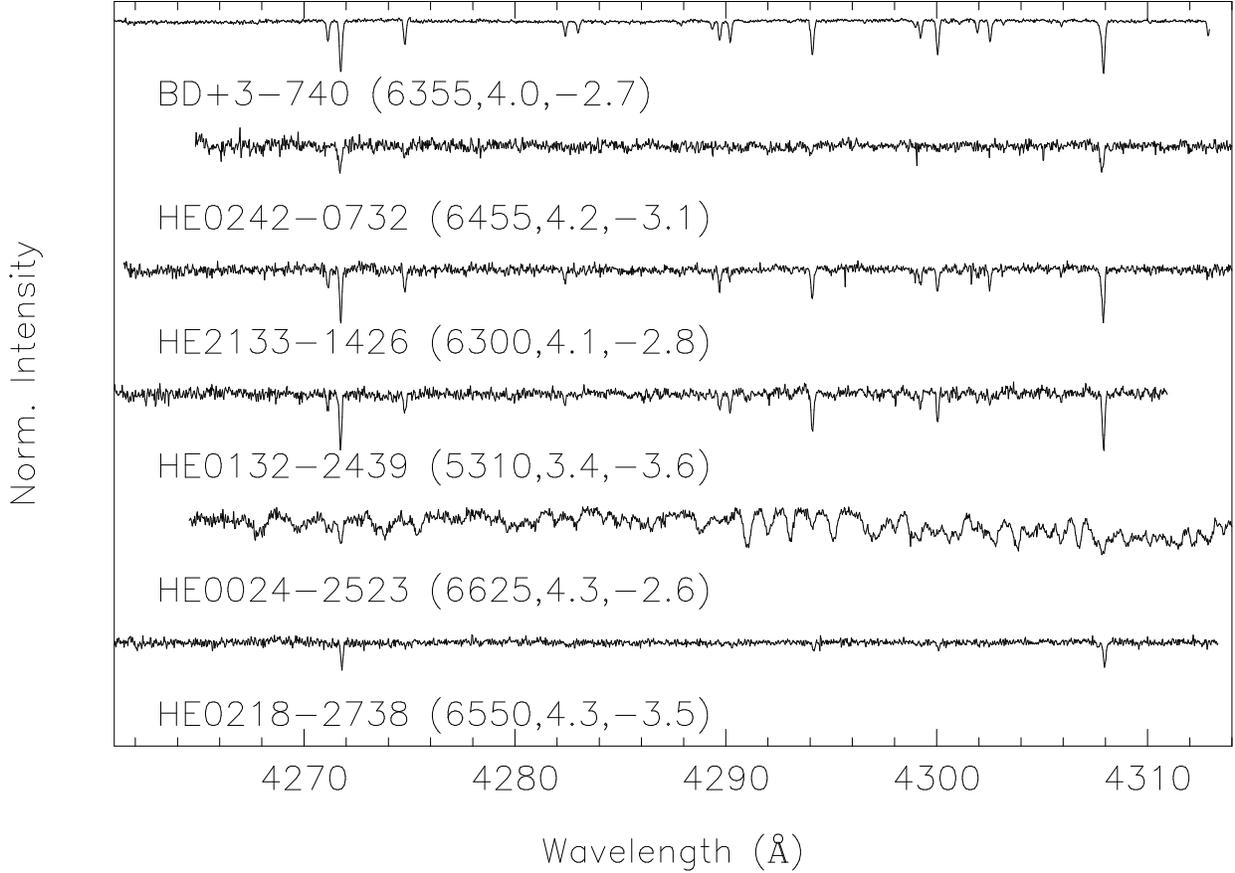}
\caption[figure1.ps]{The HIRES order containing the G band of CH
is shown for six stars.  BD+3~740 is a bright calibrating 
main sequence star; HE~0242--0732 is also a turnoff star, as is
HE~2133--1426, the faintest star in the present sample.
HE~0132--2439 is the only giant from the HES in our sample.
HE~0024--2523 is the only star which shows easily detectable CH and
its lines are resolved.
HE~0218--2738 was found to be a double-lined spectroscopic binary.
The spectra have been shifted in wavelength to remove
the effects of the difference in radial velocities among these stars.
\label{figure_6star_ch}}
\end{figure}

\begin{figure}
\epsscale{1.0}
\plotone{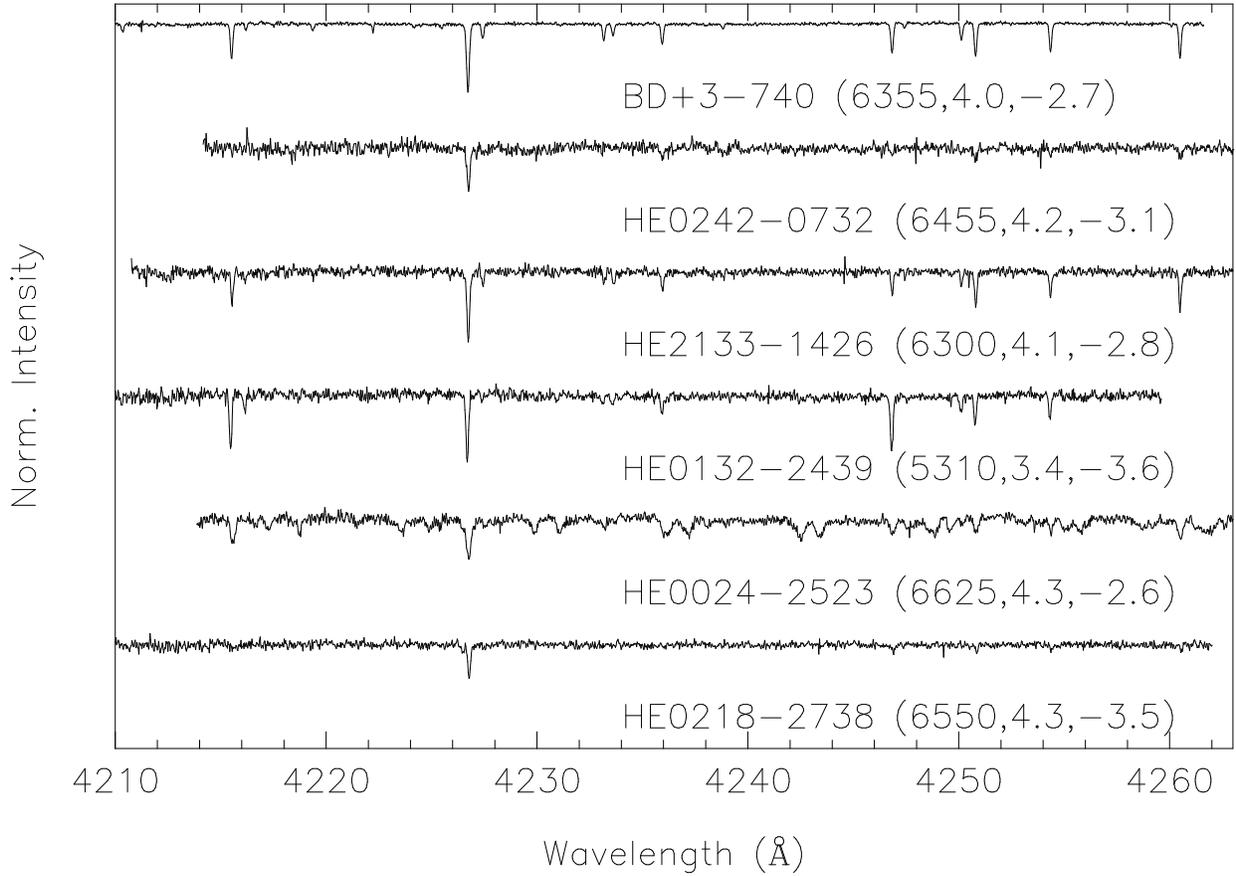}
\caption[figure1.ps]{The spectra of the same six stars are shown
in the region of the SrII line at 4215.5\,{\AA}.  The strongest line 
in this order is the CaI absorption line at 4226.7\,{\AA}.  The broader
lines found in the spectrum of HE~0024--2523 are apparent here.
The spectra have been shifted in wavelength to remove
the effects of the difference in radial velocities among these stars.
\label{figure_6star_4215}}
\end{figure}

Figures~\ref{figure_6star_ch} and \ref{figure_6star_4215} show 
sections of the
spectra for six of the stars, including the faintest star in 
our sample, a EMP main sequence turnoff star with easily
detectable absorption in the G band of CH, and a newly
discovered double-lined spectroscopic binary.  The first panel shows the order
containing the G band; the strongest two atomic absorption features
in this order are the Fe I lines at 4271.8 and 4307.9\,\AA.
In the second figure, the order containing the
Sr II line at 4215.5\,{\AA} is shown; the strongest absorption
line in that order is that of Ca I at 4226.7\,{\AA}.  To facilitate a visual
comparison of the relative strength of the Sr II line
to other absorption lines, the spectra have been
shifted in wavelength so as to remove the differences in radial velocity
between the stars displayed in the figure.  (The values of
$v_r$ are given in Table~\ref{table_specinfo}.)

The strong variation of the Sr II line relative to adjacent features
is obvious.  This variation is a well known characteristic of the spectra
of metal poor stars (e.g. Ryan, Norris \& Bessell 1991)
and will be discussed at length in Paper II.

HE~0024--2523, in addition to showing the strongest CH band, also has
absorption lines that are resolved, as is clearly shown in 
Figures~\ref{figure_6star_ch} and \ref{figure_6star_4215}.  The most
likely explanation for this is rotation.  If so, this
star has a rotational velocity of $\sim$8 \kms.  This star will
be discussed again in Paper II and also in more detail in
Gratton \etal\ (in preparation).

\section{Stellar Broad Band Photometry}

$UBV$ photometry for all the HES stars was 
obtained at the ESO-Danish 1.54\,m telescope in November, using DFOSC
(Beers \etal, in preparation).
A 1$\sigma$ error of 0.02 mag was assumed for all optical colors.
For BS 17447--029, $UBV$ was 
available from Bonifacio, Monai \& Beers (2000), but this
disagreed badly with the  $V$ photometry from 
Anthony-Twarog \etal\ (2000). Because of this large
discrepancy, $BV$ for this star
was re-observed at the 1.5-m telescope at Palomar Mountain.  The result,
given in Table~\ref{table_phot}, was identical to that of 
Anthony-Twarog \etal.  Each of the remaining two stars from the HK survey
had two independent sets of optical photometry from either
Preston \etal\ (1991), Anthony-Twarog \etal, or  
McWilliam \etal\ (1995).  These were in good agreement and the mean was used.
For the three very bright calibrating stars,
$UBV$ was taken from 
the Mermilliod, Hauck \& Mermilliod (1997) on-line data base.

We are very fortunate that uniform and reasonably precise
broad band infrared photometry at $JHK$, crucial for determination
of accurate values of \teff, is now available for all our sample stars from the
interim release of the
2MASS near infrared all-sky survey (Skrutskie \etal\ 1997). 
The global photometric
calibration of this survey is discussed in Nikolaev \etal\ (2000).  We
adopt the photometric uncertainties given in the 2MASS catalog.  
For the bright calibration stars, the near-infrared photometry is
from the most recent sources in the compilation of Gezari \etal\ (1999),
specifically Arribas \& Roger (1987),
Laird, Carney \& Latham (1988) and
Alonso, Arribas \& Martinez-Roger (1994).
The photometry is collected in Table~\ref{table_phot}.

The HES stars are bright enough that preliminary proper motions for 
most of the HES stars are now available
through the USNO CCD Astrograph Catalog (Zacharias \etal\ 2000)
for the southern sky, with
proper motions for the northern stars in the HES to follow in due course
from this same astrometric program.
Radial velocities for each EMP candidate from the HES will
be obtained as large scale vetting using  moderate-resolution 
spectroscopy
proceeds over the next few years.  Thus the full 3d velocity
vector can be specified for {\it{all}} of the HES metal poor stars.  
Such a large non-kinematically
selected sample of halo stars will be useful to compare with
increasingly sophisticated models
of the formation and evolution of the Milky Way, such as those of 
Chiba \& Beers (2001) and references therein.  A first
attempt at such a comparison has been presented by Bekki \& Chiba (2001).

Throughout this paper we use the new  Y$^2$ isochrones of Yi \etal\ (2001),
which are calculated for a scaled solar mixture.  We adopt an age of 12 Gyr,
[Fe/H] = $-$3.3 dex, with Y = 0.230, as the parameters of our standard
isochrone.  Because the metallicity of our stars are so low,
the details of the treatment of the opacities are not critical, and
there is reasonably good agreement with isochrones computed using
older stellar evolutionary codes such as those of 
Bergbusch \& VandenBerg (1992).
We will find later that the same statement holds
true for model atmospheres, for similar reasons.

\section{Interstellar Reddening}

A precise measurement of the reddening is crucial 
for the derivation of
accurate stellar parameters from broad band photometry.
Even adding the infrared does not help much if the reddening
is poorly determined in this \teff\ regime.  As is shown in
Table~\ref{table_sensitivity}, to be discussed in detail later,
the enhanced sensitivity to small changes in \teff\ one gains by
moving to optical+near-IR colors such as $V-K$ is accompanied
by an enhanced sensitivity to reddening errors.  In a
project such as mining the 
HES for EMP stars, where many stars separated by
large solid angles on the sky are involved,
control of the uniformity of the
reddening estimates is mandatory.

We adopt the extinction maps of
Schlegel, Finkbeiner \& Davis (1998)
from their analysis of the COBE/DIRBE database.
The relative extinction in various passbands is taken from
Cohen \etal\ (1981) (see also Schlegel, Finkbeiner \& Davis 1998). 

The HES includes only high galactic latitude fields
within which $E(B-V)$ is almost always $\le 0.10$ mag.  
Hence, although there is some concern that the
magnitude of the Schlegel \etal\ extinction corrections may be
slightly too large, particularly in regions where the extinction
is high ($E(B-V) > 0.15$, see, e.g. Acre \& Goodman 1999),
we adopt these values without further discussion or amendment.

All of the HES stars are fainter than $V \sim14$ as 
saturation effects within the photographic
plates used for the survey become important for brighter objects.  Given
their high galactic latitude, these
stars are sufficiently distant, even if they are main sequence
turnoff stars rather than giants, that they can all be assumed to
be beyond the reddening layer, whose thickness $H$ is only $\sim200$ pc
(see the review of Dickey \& Lockman 1990).  
Therefore the full extinction is applied to each of them.

The reddening for the brightest calibrating star (HD 140283) is
assumed to be 0.00.  This star is only about 60 pc away, and
has a Hipparcos parallax.  The other two calibrating stars
and the three stars from the HK Survey in the present sample
are not so distant as to be beyond the reddening layer, and yet
not so close that reddening can be ignored.  We therefore use
an iterative scheme
of estimating the reddening, calculating \teff\ (see below),
computing the distance $D$ 
using the luminosity from the stellar evolutionary tracks of Yi \etal\ (2001),
and then checking the estimate of the reddening
against the quantity $E(B-V) \times e^{-D sin(b)/H}$.  Similar
schemes are described by Laird, Carney \& Latham (1988) and by
Bonifacio \etal\ (2000).
Rapid convergence for $E(B-V)$ is obtained.

The recommended values of $E(B-V)$ are given in the last column of
Table~\ref{table_phot}.  The reddenings for the HES stars are quite small,
with all but one of our sample having $E(B-V) < 0.03$ mag.
It is only for the brighter calibration objects that this issue
becomes a major concern.

\section{\teff\ from Broad Band Colors \label{text_teff}}

Because of concerns about potential non-LTE effects, which are suspected to be
stronger in very metal poor stars than in stars of solar metallicity, 
we must derive \teff\
from broad band colors.  Detailed calculations for non-LTE in
Fe are presented in Th\'evenin \& Idiart (1999) and in
Gratton \etal\ (1999).  Although the most careful analyses of
globular cluster and field
stars of higher metallicity than those considered here, such as that
of  Cohen, Behr \& Briley (2001) and 
Ram\'{\i}rez \etal\ (2001) for a large sample of stars
over a wide range in luminosity in M71, Ivans \etal\ (2001)
for a large sample of red giants in M5,
or Allende Prieto, Asplund, Garcia Lopez \& Lambert (2002) for Procyon,
show that departures from LTE in the formation of Fe lines are
relatively small, the predicted strength of the departures
from non-LTE increases as the metallicity decreases.  To be 
conservative, at least initially,
measurements based on ionization equilibrium
in the spectra themselves cannot be considered as reliable at this time.

We utilize here the grid of predicted broad band colors and
bolometric corrections of
Houdashelt, Bell \& Sweigart (2000) based on the
MARCS stellar atmosphere code (Gustafsson \etal\ 1975). 
We assume that their Johnson-Glass $J,K$ colors are
equivalent to the $J,K_s$ 2MASS colors. 
Carpenter (2001) has derived the transformations between the 2MASS 
photometric system and many other infrared photometric systems, and
finds a very small zero point offset between these two infrared
photometric systems, which we have chosen to ignore until the
final release of the 2MASS catalog becomes available.
 
Cohen, Behr \& Briley (2001) have demonstrated 
that the Kurucz and MARCS predicted  $V-K$ colors are essentially
identical, at least for $V-K$ colors and for the set of
models with [Fe/H] = $-$0.5 dex.
 
Since we are interested in EMP stars, the details of the
treatment of the opacity from any element other
than H should not be important, and we expect to continue to find
very good agreement for the predicted broad band colors
from the various grids of model
atmospheres available.  In our tests,
we take the predicted $V-K$ color
from each model in the MARCS grid with [Fe/H] = $-3.0$, and interpolate
within the Kurucz color grid at the same abundance and at the \grav\
of the MARCS model
to find  the \teff\ that would be deduced.
We continue this comparison, checking $V-K$, $V-J$ and $B-V$
at [Fe/H] = $-$3.0 dex.  We find that
the colors predicted from the MARCS code from Houdashelt \etal\
are essentially identical to those from 
the Kurucz ATLAS code (Kurucz 1992) at this low metallicity. 

\begin{figure}
\epsscale{1.0}
\plotone{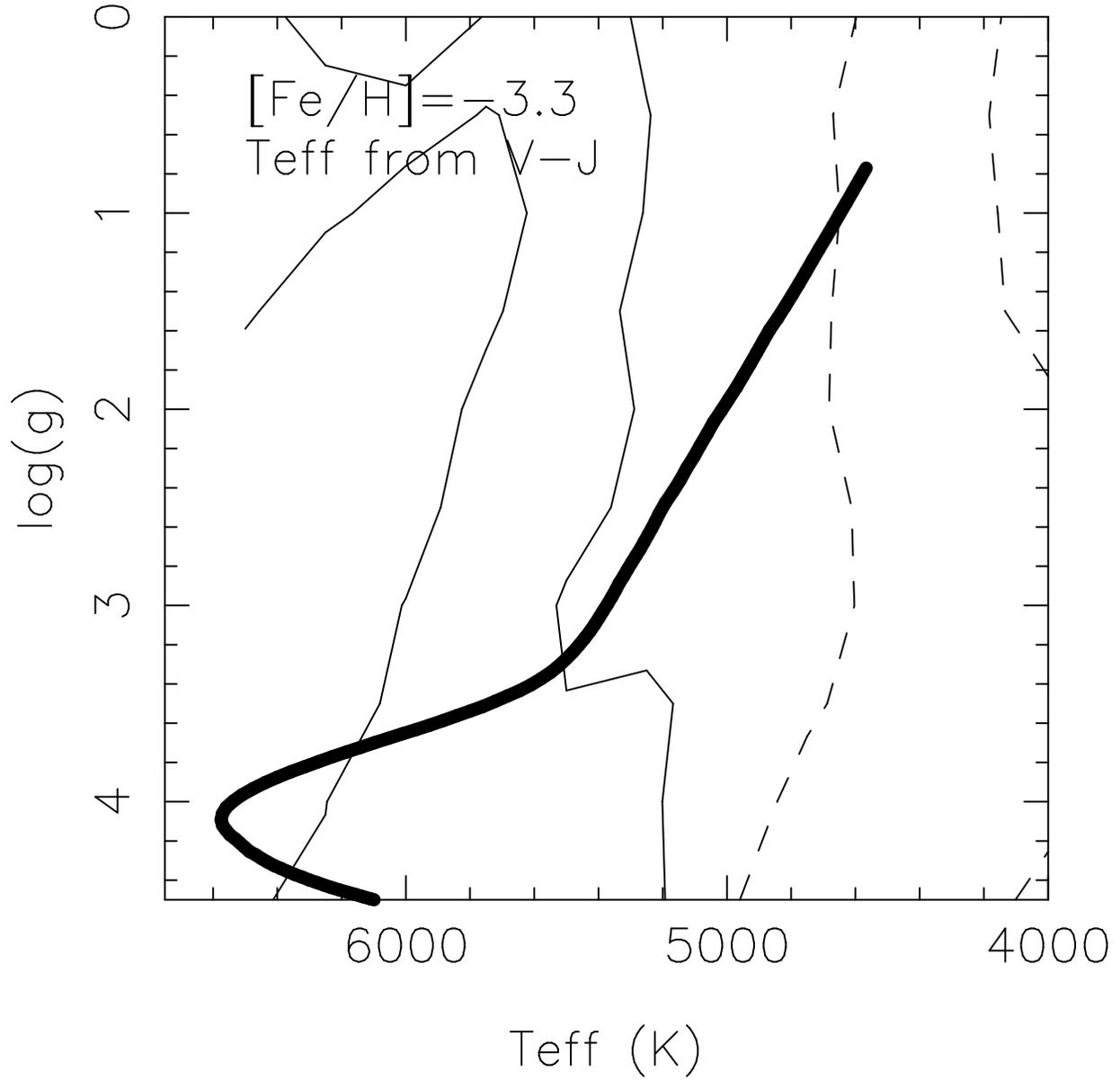}
\caption[figure1.ps]{A comparison of the Kurucz and MARCS
temperature scale from $V-J$ colors.  The four contour levels shown
correspond to $\Delta$\teff = $-30$, $-10$, 10 and 30 K. 
The thick curve is a 12 Gyr isochrone for a very metal poor star
from  the very recently completed Y$^2$ grid of isochrones of
Yi \etal\ (2001). 
\label{figure_vj_comp}}
\end{figure}

A contour plot of the difference $\Delta$\teff(Kurucz - MARCS) 
that
results when the $V-J$ color is used
is shown in Figure~\ref{figure_vj_comp}.  The four contour levels displayed
correspond to $\Delta$\teff = $-30$, $-10$, 10 and 30 K.  Also shown
in this figure as the thick curve is a 12 Gyr isochrone for Fe/H = $-2.3$ dex
from  the very recently completed Y$^2$ isochrones of
Yi \etal\ (2001).
Along this isochrone, $\Delta$\teff = 0 to 30 K
for the subgiants and main sequence turnoff region, and $\Delta$\teff = 
0 to $-30$ K for the red giant branch.    We thus demonstrate that,
to within a tolerance of $\pm30$ K, the Kurucz and MARCS temperature scales
from broad band $V-J$ colors are identical.

We therefore proceed to assign values of \teff\ using the grid of colors
of Houdashelt \etal\ (2000).  For the calibrating stars and for
some of the brighter stars from the HK survey, 
a high dispersion abundance analysis has already been done.  These
are summarized in  Table~\ref{table_oldabund}.  In these cases we
take the [Fe/H] from such an analysis as our initial guess for
the stellar metallicity.  If not, we assume
[Fe/H] = $-3.0$ dex.  We adopt initial guesses for log(g) based
on whether the star is a giant or a main sequence turnoff star.
These rough guesses are used to interpolate to the proper value
in the desired color, which depends slightly on \grav\ and abundance,
as well as on \teff.  

\begin{figure}
\epsscale{1.0}
\plotone{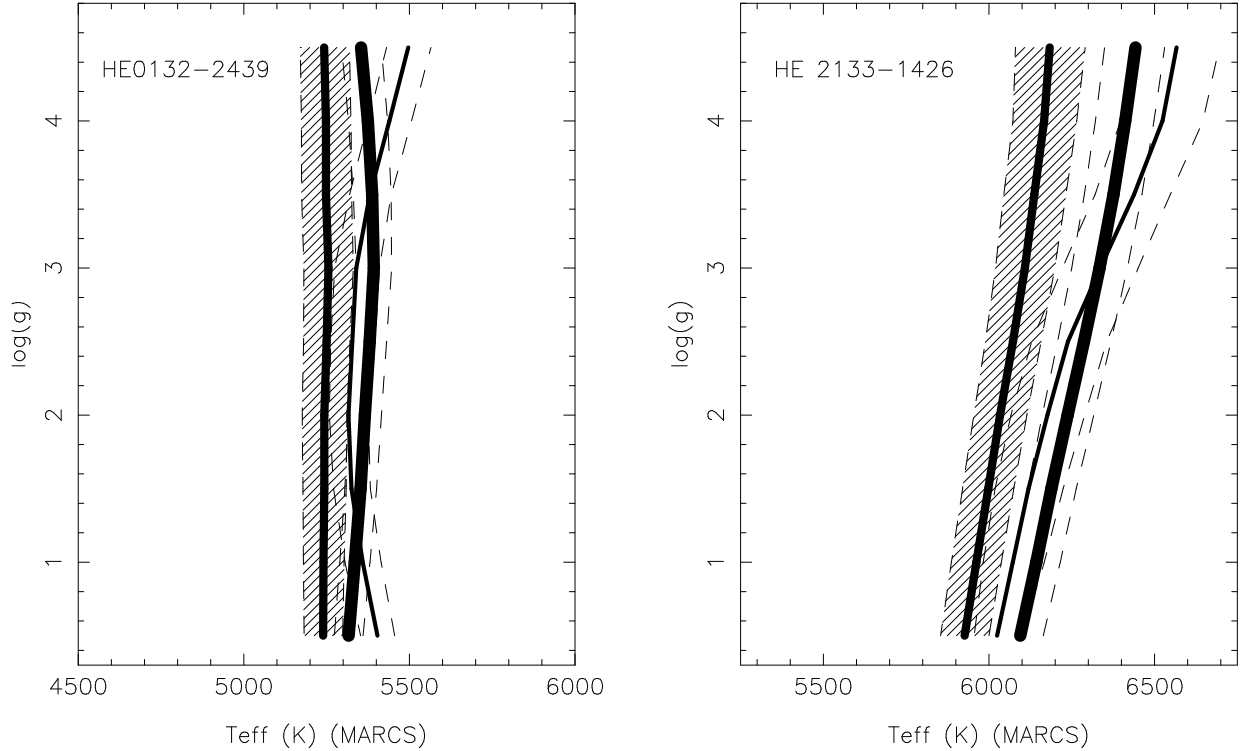}
\caption[figure2.ps]{Examples of our procedure for deriving
\teff\ from stars in our sample are shown. 
The left panel is for the star HE~0132--2439,
an EMP giant. The solution for its \teff\
from its dereddened colors
is displayed as a function of \grav\  with [Fe/H] assumed
to be $-3.0$ dex.  The results from the $B-V$ measurement 
are shown as the thinnest solid curve,
from the $V-J$ color as thicker solid curve, and from the 
$V-K$ color as the thickest solid curve.  Dashed curves 
denote the values of \teff\ inferred from the measured and
dereddened colors $\pm$ their 1$\sigma$
uncertainties.  The striped area indicates that allowed 
within this 1$\sigma$ uncertainty 
level inferred from the $V-J$ color.
In the second panel, we show the \teff\ determination for the faintest
star in the present sample, HE~2133--1426, a EMP star near the
main sequence turnoff. 
\label{figure_phot_match}}
\end{figure}

This process is illustrated for a subgiant and for a main sequence
turnoff region star (the faintest star in the present sample)
in Figure~\ref{figure_phot_match}.
The results for three colors, $B-V$, $V-J$ and $V-K$, are shown
as solid curves, with the thickness of the curves varying,
$B-V$ having the thinnest and $V-K$ having the thickest curves.
Dashed curves denote the \teff\ deduced for the observed and dereddened colors
$\pm$ the 1$\sigma$ uncertainty of each observation.
In a perfect world, all the solid curves would
overlap to within the uncertainties at some definite
value of \teff, and this does in fact occur for the stars shown
in the figure.  
We have found that the $B-V$ colors give systematically higher
\teff, by about 50 K, for the main sequence EMP stars (see 
Figure~\ref{figure_phot_match}).

Table~\ref{table_sensitivity} gives the sensitivity of each color to
various stellar parameters.  These are expressed as the change in
the parameter (which may be positive or negative)
that would result from a $+1\sigma$ change in the color,
where the observational uncertainty for each color defines 
the value of $\sigma$ used.  These calculations were made using
stellar parameters appropriate for EMP main sequence turnoff stars.
We infer from this table, for example, that $B-V$ is
less sensitive to changes in \teff\ than either $V-J$ or $V-K$,
and that none of the colors considered is capable of applying
a significant constraint on the metallicity of a star.

The values of \teff\ that result from this procedure,
applied to each star in our sample, are listed in Table~\ref{table_param}.
As one might expect from Table~\ref{table_sensitivity},
they have typical internal errors of $\pm$75 K, except for those
from $B-V$, where the error is $\sim\pm$125 K.

$U-B$ colors are also available, but they are more difficult to match
to the predictions of the model atmosphere grid.  
For five of the eight stars from the HES, $U-B < -0.25$ mag when
corrected for reddening.  This is very blue, so much so that such
a blue color is not reached in the relevant regime of \teff\ and 
\grav, even at [Fe/H] = $-3.0$ dex, the most metal poor models
in the grid of Houdashelt \etal\ (2000). In addition, the predictions for $U-B$ depend sensitively on
metallicity as well as \teff\ and \grav, and are non-monotonic
in these two parameters.

While the contribution from line opacities becomes less important 
in defining continuum fluxes at the lowest metallicities,
some problems in the violet and ultraviolet may remain.
We believe that these problems
of properly predicting the flux in the region 3500 -- 4500\,{\AA}
in these EMP stars are related
to the extremely low metallicity of these stars, to the 
sensitivity to multiple parameters, and to the relatively
small sensitivity to \teff. 
This situation is alleviated
by using redder colors with wider wavelength separation.
Therefore the adopted \teff\ given in Table~\ref{table_param}
are the means of those
deduced from the dereddened $V-J$ and the $V-K$ colors.

We have also determined \teff\ through examination of the Balmer line
profiles. H$\alpha$ is the Balmer line best suited for this purpose, because
it has the strongest temperature sensitivity, and its profile is almost
independent of the choice of the mixing-length parameter $\alpha$ (e.g.,
Fuhrmann \etal\ 1993).  Unfortunately, H$\alpha$ is not covered by the
spectral range chosen in this work.  H$\beta$ is not well centered in the
echelle orders, hence unsuitable.  The continuum near H$\gamma$ may be
perturbed by strong CH lines in carbon-enhanced stars, which occur at rate of
$\sim 20$\,\% among EMP stars. Hence we utilize
H$\delta$ for this purpose.  A special reduction of the relevant order was
made to ensure the best possible continuum level.  The orders above and below
this were used as necessary to interpolate the continuum level across the
Balmer line.  The synthetic Balmer line profiles computed by the Gehren group
(Fuhrmann 1998, 2000 private communication) are used. The results 
are listed in the fifth column of
Table~\ref{table_param}. These have a typical error of $\pm100$ K.

\begin{figure}
\epsscale{1.0}
\plotone{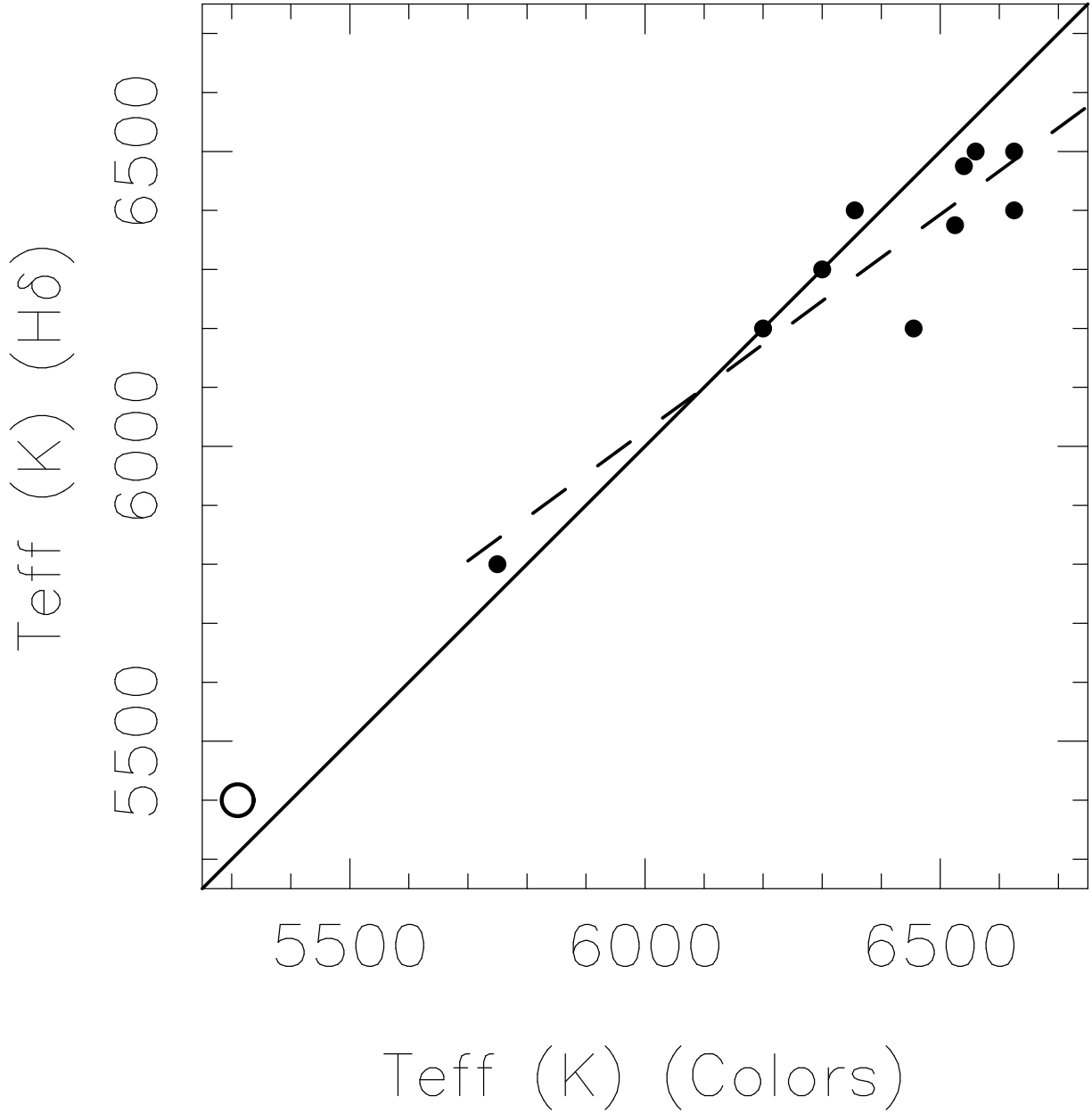}
\caption[figure3.ps]{The \teff\ values deduced from the analysis
of the profiles of the H$\delta$ lines in our sample of very metal poor
stars is shown as a function of the \teff\ values derived from the
observed, dereddened colors.   The solid line denotes equality, while
the dashed line indicates the best linear fit.
The large open circle denotes the only subgiant in our present sample,
which is not included in the fit. 
The $\sigma$ about the best fit line is $\sim$75 K. 
\label{figure_balmer}}
\end{figure}

A perusal of Table~\ref{table_param} shows that we have achieved
consistency to within $\pm100 K$ between the \teff\ from 
the Balmer line profile and
from the broad band colors, as is also shown in Figure~\ref{figure_balmer}.
The solid line denotes equality between the \teff\ from the
broad band photometry and from the H$\delta$ profiles, while the
dashed line is the best linear fit excluding the single subgiant,
which suggests that the \teff\
for the turnoff stars from the broad
band photometry is systematically slightly hotter than that inferred from
H$\delta$.

We note
that the \teff\ we assign to these main sequence turnoff stars is 
identical to that found through stellar evolution, i.e. the predicted
location of the main sequence turnoff in 12 Gyr isochrones for
metal poor stars, which is 
at about 6600 K for [Fe/H] = $-2.3$ and about 6750 K for [Fe/H] = $-3.3$ dex. 
If the \teff\ we assign are consistent with the \teff\ scale adopted
for the Y$^2$ grid of isochrones, ages older than 14 Gyr can be ruled
as the main sequence turnoff then becomes cooler than the 
observed values reached by these stars. Such old ages are also
not supported by recent Boomerang observations of the fluctuations in
the CMB (de Bernardis \etal\ 2000).  

\subsection{Comparison With the Empirical \teff\ Scale of Alonso \etal}

We have compared our \teff\ scale with the
empirical color--\teff--[Fe/H] relations for dwarfs and for
giants established by Alonso \etal\ (1996, 1999).
We use their polynomial fit for the $V-K$ colors, simulating moving
down in luminosity along an isochrone from the tip of the RGB to the main sequence
for EMP stars.  Since the Alonso \etal\ fits are not well
calibrated at extremely low metallicities, we use their
fits with [Fe/H] = $-2.0$ dex; the metallicity sensitivity of
$V-K$ at lower abundances than this is small.

We find that \teff\ inferred from the Alonso \etal\ relations 
for the $V-K$ color is 
150 K cooler than that inferred from the Houdashelt \etal\ (2000) grid
of colors predicted from stellar atmosphere models
near the top of  the giant branch, but this decreases
to 100 K cooler for subgiants and for stars
near the main sequence turnoff for [Fe/H] = $-2.0$ dex.  
Pushing the Alonso \etal\ relations beyond their regime
of validity and verification to [Fe/H] = $-3.0$ dex would
produce good agreement at the main sequence but the subgiants and
giants would still be 100 K hotter using the MARCS color 
grid.\footnote{This may be related to problems encountered 
by Lebreton (2000) in matching
stellar evolutionary isochrones with stellar parameters for nearby
stars with Hipparcos parallaxes.}
The Alonso scale thus is close to
the \teff\ determined from H$\delta$.

Thus, at least with respect to the $V-K$ color, 
the Alonso \etal\ scale and our \teff\
scale are in good agreement for turnoff stars.  The small
systematic difference of a maximum of 100 K for giants would
translate into a change in derived metallicity of 
$\Delta$[Fe/H] $\sim -0.1$ dex.

Because the Alonso \etal\ empirical \teff\ scale is not calibrated
for stars as metal poor as our sample,
and because of our philosophical preference to 
utilize the same model atmospheres that one must rely upon for
analyses of the spectral features, we have chosen to adopt
the \teff\ scale established from broad band optical-IR colors.
One potential systematic error for this choice is the 
matching of the photometric systems between the observational colors
and the theoretically predicted ones.
We believe that the internal
errors in our adopted \teff\ are $\pm$75 K, with a possible systematic 
overestimate (arising from as yet unknown causes) 
for the main sequence turnoff stars only 
of $\sim$100 K. 

\subsection{Comparison With Previous \teff\ Determinations}

We compare our values of \teff\ with the limited set of values available
from previous analyses of the brighter stars in our sample.  We adjust
the \teff\ values listed in Table~\ref{table_param} for the
difference in reddening between the values we adopt given in 
Table~\ref{table_phot} and those adopted in previous analyses.
The results are given in Table~\ref{table_tcompare}. 
Once the reddenings are forced to the same value, each
of the five entries has 
${|}\Delta($\teff)$({\rm{old-now}}){|} < 100$ K.
This is very encouraging.

\section{Surface Gravities}

The only broad band color among those available to us
with strong sensitivity to \grav\ is
$U-B$.  However, for the reasons described in \S\ref{text_teff},
we choose not to use it except in the crudest possible sense,
i.e. to discriminate among turnoff stars and subgiants and giants.
We cannot utilize the usual spectroscopic indicator, the ionization
equilibrium, because of the possibility of non-LTE effects as
described above.

Once \teff\ is known,
we determine \grav\ from the Y$^2$ isochrone for stars
that are 12 Gyr old with Z = $1.0 \times 10^{-5}$,
equivalent to [Fe/H]=$-3.3$.  The \grav\ is determined directly
from the isochrone using the \teff\ of the star. 
If the 14 Gyr isochrone of the
same metallicity from the Y$^2$ grid were adopted, the assigned \grav\ would
increase by an amount which is negligible for the main sequence
stars and which increases as \teff\ decreases, to 0.15 dex for \teff\ = 5000 K.

This procedure is not very precise
for giants because the slope of the RGB is rather steep,
$\Delta$(\grav/\teff) = $\sim$0.3 dex/100 K.  Consideration of the 
rate of evolution along the isochrone as a function
of luminosity, which directly translates into the number density of stars
along the isochrone, dictates that the probability of finding
subgiants is considerably higher in faint samples such as the
HES than that of finding stars
well up the RGB. 

For stars near the main sequence turnoff, which is the location of the
bulk of the present sample, the slope of the
isochrone is much shallower and good discrimination in \grav\
is possible.  

However, there are two solutions in the isochrone for a fixed \teff\ near the
main sequence turnoff, one corresponding to main sequence stars slightly
fainter than the turnoff, and one to stars which have just started to evolve
off the main sequence toward the base of the giant branch; these two solutions
differ by about 1 dex in $\log g$\ for \teff$\sim 5900$~K. For \teff\ more
than $\sim$500 K cooler than the turnoff itself, the lower luminosity of the
main sequence stars makes their detection quite unlikely, even though their
volume density is higher than the post-turnoff main sequence stars. In fact,
assuming a constant volume density over the galactic region sampled by HES
\footnote{Typical distances of turnoff-stars in the HES survey are $\sim 1$~kpc;
this is much less than the local value
of the halo scale height.}, a comparison with globular
cluster luminosity functions show that at this temperature we expect to detect
approximately one main sequence star out of 10 to 15 subgiants. For \teff\
closer to the turnoff, both solutions have fairly similar 
probabilities, although
the brighter (lower gravity) solution is always more probable.

We cannot distinguish between these two cases using our present photometric 
dataset alone.  Making this distinction using the Fe ionization
equilibrium (i.e. relying on the spectra, and on the 
understanding of or absence of
large non-LTE effects) is only feasible in the best of our spectra
as the difference in \grav\ between the two solutions
is not large.  As noted in Table~\ref{table_param},
the detailed analysis of the spectra presented in Paper II suggests
that 
two of the three bright comparison stars near the main sequence turnoff
actually are slightly fainter than the turnoff.  All of
the HES sample members have been assigned to be brighter than the
main sequence turnoff.

Only one of the stars among the bright calibration objects has
an accurate Hipparcos parallax, HD 140283, discussed by Korn \& Gehren (2000).
We adopt their value of \grav.  Our derived \grav\ for
BD+3~740 is consistent with that inferred from the rather uncertain
Hipparcos parallax for that star (7.8$\pm$2.1 mas), 
assuming a mass of 0.8 M\subsun.

We emphasize that our \teff\ determination is independent of any
spectroscopic criteria, and that the choice of \grav\ is made
based on \teff\ and an assumed isochrone.  It is only when there
are multiple possible values of \grav\ with approximately
equal probability, which occurs only near the main sequence
turnoff, that spectroscopic criteria
are used.  The difference in \grav\ between the two solutions
does not exceed 0.5 dex before the lower luminosity solution
becomes highly improbable.   The details of abundance analyses to
be presented in Paper II will verify the validity of this process.

\section{Cautionary Notes}

Kurucz (2001) has reviewed the many deficiencies of the current
state of stellar astrophysics.  The treatment of non-LTE
in essentially all abundance analyses  may
be inadequate.   (We only apply non-LTE corrections for
a selected small set of ions.  See Paper II.) 
Our treatment of convection surely could be
better. Detailed three dimensional 
radiative-hydrodynamic convection simulations
for solar granulation such as those of Asplund \etal\ (2000)
now exist.  The first steps towards
utilizing such models (Allende-Prieto \etal\ 2002) to study line profiles and
abundances
are now starting to become available, although 
such an elegant treatment still requires large 
amounts of computing time and very high precision spectra, with much
better spectral resolution and SNR than those discussed here.

In our present application, the question is how well does the continuum and
its broad band colors, or the Balmer-line profiles, represent the temperature
gradient within the atmosphere.  This is what is explored by the absorption
features, as strong lines tend to arise from levels with smaller 
excitation potentials,
and hence are formed farther out in the atmosphere.
Kurucz points out that Balmer line wings
will be preferentially formed in the hotter convective elements
due to the high excitation potential of the second level of HI.
Hence a simple one dimensional model such as we are using will
have a somewhat higher \teff\ deduced from the Balmer lines
than does the real star.
However, with such metal poor stars, all the absorption lines
are very weak, and the issue of the temperature gradient within
the atmosphere is perhaps less critical, as the lines are formed closer
to the mean depth of formation of the continuum.

\teff\ obtained from the excitation temperature
of Fe should be reliable, and disagreements
between these atmospheric parameters and those from $T_{exc}$
might indicate problems with the correct temperature stratification
in the atmosphere.  Such problems have been noted in the past by
Dalle Ore (1992) and by Gratton, Carretta \& Castelli (1996), with a more
limited discussion given in Fulbright (2000).
These problems appear to be considerably worse in the very metal
poor giants than in the dwarfs.  This issue
will be taken up again in the following paper.

We have adopted a prescription, based on
photometric indices for determining the stellar parameters of
our very metal poor stars, that is quite different from that used
in most previous analyses, which rely much more on spectroscopic
equilibria and/or exclusively on bluer colors.  By rejecting
the spectroscopic equilibria in choosing \teff, and by
using theoretical isochrones to determine \grav, we leave open the possibility
that our derived Fe abundances may have a dependence on the lower
excitation potential of the transition, $\chi$.  This in fact
turns out to be the case; we do find a small dependence
in some stars in Paper II.  The possible existence of this dependence
means that care must be exercised in comparing our abundances
with those from other analyses, as the distribution of line
excitation potentials in the two samples must be taken into 
account.

\section{The Yield of the HES for EMP Stars}

In the last column of Table~\ref{table_specinfo} we give the 
Fe abundance ([Fe/H] deduced from Fe I lines) for the present
sample, taken from Paper II.  For the bright stars with previous high dispersion
abundance analyses, listed in Table~\ref{table_oldabund}, we find
that our abundances are comparable when the \teff\ values
adopted in both analyses are essentially identical 
(see Table~\ref{table_tcompare}), and
tend to be slightly higher in those cases
where our adopted \teff\ is hotter, as expected.

Three of the eight HES candidate EMP stars 
in the present sample do in fact have [Fe/H] $\le -3.0$ dex,
with three others having [Fe/H] $\le -2.8$ dex.  The sample of two very metal
poor stars observed by Depagne \etal\ (2001) at the VLT also yielded
metallicities close to those expected from their moderate-resolution
follow up spectra.
This means that the process of
automatic selection of EMP candidates from the HES, followed by
a visual check, then
further vetting using moderate-resolution spectroscopy, produces
samples of EMP stars with a high yield.

It is interesting to compare our derived [Fe/H] values from the
present set of HIRES spectra with the abundance inferred for the
stars in our sample from the moderate-resolution follow up
spectroscopy. Those values are deduced from the strengths of the
Ca II absorption at 3933\,{\AA}, the Balmer lines, and the $B-V$ colors
with the standard method developed for the HK survey, and most recently
described in Beers \etal\ (1999)
\footnote{Some additional features, such as auto-correlation
function measurements, were added to the
standard abundance estimation algorithms for the HK
Survey by Beers \etal\ (1999); those were not used for the HES stars.},
and are given in the next to last column of 
Table~\ref{table_specinfo}, together with a $1\sigma$ error on the
abundance determination.  The metallicities obtained from the
moderate-resolution spectra are, in general, somewhat more metal-poor,
part of which may arise from the somewhat hotter \teff\ values
adopted here.  The
difference in the mean [Fe/H]$_K$ - [Fe/H]$_{HIRES}$ is $-0.23$ dex, with a
dispersion about this mean offset of 0.32 dex.  This dispersion is in
accordance with expectation, given the listed errors in [Fe/H]$_K$  and the
smaller errors expected for the present effort.  A detailed
discussion of these moderate-resolution spectra and the derivation
of metallicities therefrom will be given elsewhere (Christlieb \etal\,
in preparation).

All of this is very
encouraging for future large scale projects which plan to mine the HES
database for extremely metal poor stars.

\section{Summary}

In this paper and its companion paper, which presents a detailed
abundance analysis for these HIRES spectra,
we have demonstrated
the high effective yield of the HES for extremely metal
poor stars, and we now
intend to embark on a major project (the 0Z Project) 
to understand the early chemical
evolution of the galactic halo.  

Our ultimate goal, which will take perhaps
five years to reach with several large observatories working in a coordinated
fashion, is high dispersion abundance analyses of a sample of 500
extremely metal poor stars.  We will gain a detailed
knowledge of chemical evolution during the initial phases of the
formation of the Galactic halo, the kinematic properties of the
extremely metal poor halo stars, a much better knowledge of the
metallicity distribution of very metal poor halo stars, a much
better estimate for the metallicity of the most metal poor stars
(currently [Fe/H] $=-4.0$ dex).  A detailed study of those
extremely rare r-process enhanced stars that may be found
similar to that of Cayrel \etal\ (2001)
will improve our age estimates for the Galactic halo 
based on
radioactive decays of unstable heavy elements.

\acknowledgements
The entire Keck/HIRES user community owes a huge debt
to Jerry Nelson, Gerry Smith, Steve Vogt, and many other people who have
worked to make the Keck Telescope and HIRES a reality and to
operate and maintain the Keck Observatory.  We are grateful
to the W. M. Keck Foundation for the vision to fund the construction 
of the W. M. Keck Observatory.  The authors wish 
to extend
special thanks to those of Hawaiian ancestry on whose sacred mountain
we are privileged to be guests.  Without their generous hospitality,
none of the observations presented herein would
have been possible.

This publication makes use of data products from the Two Micron All Sky
Survey, which is a joint project of the University of Massachusetts and
the Infrared Processing and Analysis Center/California Institute
of Technology, funded by the National Aeronautics and Space Administration
and the National Science Foundation.  This research has made use of the SIMBAD
database, operated at CDS, Strasbourg, France.

We thank T. Gehren and J. Reetz for providing us with Balmer line
      profiles, and making the spectrum investigation utility SIU available
      to us.  We thank the referee, Bruce Carney, for his thoughtful and
constructive comments.
We thank Sarah Yost, James Chakan and Hubert Chen 
for obtaining the $BV$ photometry of BS 17447--0029. 

T.C.B acknowledges partial support for this work from grants AST 00-98508 and
AST 00-98549 from the National Science Foundation.

\clearpage

%
%
\begin{deluxetable}{lrrr}
\tablenum{1}
\tablewidth{0pt}
\scriptsize
\tablecaption{Coordinates for the HES Stars in the HIRES Sample} 
\tablehead{
\colhead{Name} & \colhead{$V$} & \colhead{RA} & \colhead{Dec} \nl
\colhead{} & \colhead{(mag)} & \colhead{(J2000)} & \colhead{(J2000)} \nl
}
\startdata
HE~2133--1426 & 15.484 & 21 36 07.2 & $-$14 12 36 \nl
HE~2344--2800\tablenotemark{a} & 14.856 & 23 46 44.4 & $-$27 44 10 \nl
HE~0024--2523\tablenotemark{b} & 14.913 &  00 27 27.6 & $-$25 06 26 \nl
HE~0130--2303 & 14.758 & 01 33 18.2 & $-$22 48 36 \nl
HE~0132--2439 & 14.821 & 01 34 58.8 & $-$24 24 18 \nl
HE~0148--2611 & 14.453 & 01 50 59.5 & $-$25 57 02 \nl
HE~0218--2738\tablenotemark{c} & 14.883 & 02 21 04.0 & $-$27 24 40 \nl
HE~0242--0732 & 15.793 & 02 45 00.6 & $-$07 19 42 \nl
\enddata
\tablenotetext{a}{This is a re-discovery of CS 22966--048, originally
found in the HK survey.}
\tablenotetext{b}{This star has easily detectable CH absorption in the G band
and its lines are resolved.}
\tablenotetext{c}{The HIRES spectra show this star is a 
double-lined spectroscopic binary.}
\label{table_coords}
\end{deluxetable}

%
%
\begin{deluxetable}{l r rrrr rrr}
\tablenum{2}
\tablewidth{0pt}
\scriptsize
\tablecaption{Observed Colors and Details of the HIRES 
Spectra for the Stellar Sample} 
\tablehead{
\colhead{Name} & \colhead{$V$} & \colhead{$U-B$} & 
\colhead{$B-V$} & \colhead{$J$} & \colhead{$K$} & 
\colhead{Exp. Time} & \colhead{SNR\tablenotemark{a}} & \colhead{$E(B-V)$} \nl
\colhead{} & \colhead{(mag)} & \colhead{(mag)} & \colhead{(mag)} &
\colhead{(mag)} & \colhead{(mag)} & \colhead{(sec)} &
\colhead{(at 4625\,{\AA})} & \colhead{(mag)} \nl
}
\startdata
Very Bright \nl
HD 140283 & 7.211 & $-$0.196 & 0.490 & 6.03~ & 5.63~ & 75 & 300 & 0.000 \nl
BD+3~740 & 9.808 & $-$0.20~ & 0.36~ & 8.77~ & 8.49~ & 240 & 240 & 0.037 \nl
G139--8 & 11.508 & $-$0.203 & 0.475 & 10.345 & 10.031 & 650 & 165 & 0.067 \nl
  &  \nl
HK Stars \nl
BS 17447--029 & 13.57~ & $-$0.30~ & 0.40~ & 12.583 & 12.313 & 1800 & 113 
      & 0.047 \nl
CS 22878--101 & 13.78~ & 0.299 & 0.799 & 11.87~ & 11.26~ & 3600 & 96
      & 0.086 \nl
CS 22950--046 & 14.224 & 0.345 & 0.906 & 12.228 & 11.570 & 2400 & 93
      & 0.108 \nl
  &  \nl
HES Stars \nl
HE~2133--1426 & 15.484 & $-$0.237 & 0.449 & 14.374 & 14.120 & 9900 & 105
     & 0.052 \nl
HE~2344--2800\tablenotemark{b} & 14.856 & $-$0.191 & 0.388 & 13.930 
      & 13.763 & 9600 & 125
     & 0.018 \nl
HE~0024--2523 & 14.913 & $-$0.254 & 0.408 & 14.067 & 13.740 & 6000 & 98
     & 0.017 \nl
HE~0130--2303 & 14.758 & $-$0.225 & 0.393 & 13.901 & 13.554 & 6000 & 120
     & 0.012 \nl
HE~0132--2439 & 14.821 & $-$0.117 & 0.630 & 13.379 & 12.931 & 4800 & 90
     & 0.012 \nl
HE~0148--2611 & 14.453 & $-$0.250 & 0.371 & 13.545 & 13.291 & 3600 & 102
     & 0.013 \nl
HE~0218--2738\tablenotemark{c} & 14.883 & $-$0.240 & 0.394 & 13.979 & 13.711 & 7200 & 115
     & 0.013 \nl
HE~0242--0732 & 15.793 & $-$0.271 & 0.434 & 14.795 & 14.567 & 9300  & 94 
      & 0.027 \nl
\enddata
\tablenotetext{a}{SNR is given per spectral resolution element (3 pixels).}
\tablenotetext{b}{This is a re-discovery of CS 22966--048, originally
found in the HK survey.}
\tablenotetext{c}{The HIRES spectra show this star is a 
double-lined spectroscopic binary.}
\label{table_phot}
\end{deluxetable}

%
%
\begin{deluxetable}{lrllrr}
\tablenum{3}
\tablewidth{0pt}
\scriptsize
\tablecaption{Recent High Dispersion Abundance Analyses of The 
Brighter Sample Stars} 
\tablehead{
\colhead{Name} & \colhead{\teff} & \colhead{\grav} & 
\colhead{[Fe/H]} & \colhead{$E(B-V)$} &\colhead{Ref.}  \nl
\colhead{} & \colhead{(K)} & \colhead{(dex)} & \colhead{(dex)} &
\colhead{(mag)} & \colhead{}  \nl
}
\startdata
Very Bright \nl
HD 140283 & 5750 & 3.4 & $-$2.54 & 0.01 & a \nl
          & 5680 & 3.5 & $-2.64$ & 0.00 & b \nl
          & 5810 & 3.67 & $-$2.29 & ...\tablenotemark{h} & c \nl
          & 5650 & 3.4 & $-2.4$ & ...\tablenotemark{i} & d \nl
BD+3~740 & 6240 & 4\tablenotemark{j} & $-$2.70 & 0.01 & e \nl
          & 6075 & 3.8 & $-2.7$ & ...\tablenotemark{i} & d \nl
G139--8 & 6025 & 4\tablenotemark{j} & $-$2.56 & 0.04 & f  \nl
  &  \nl
HK Stars \nl
CS 22878--101 & 4780 & 1.15 & $-3.13$ & 0.06 & g \nl
CS 22950--046 & 4635 & 0.85 & $-3.40$ & 0.06 & g \nl
\enddata
\tablenotetext{a}{Ryan, Norris \& Beers (1996)}
\tablenotetext{b}{Nissen \etal\ (1994) and Edvardsson \etal\ (1994)}
\tablenotetext{c}{Korn \& Gehren (2000) -- using \grav\ from 
Hipparcos parallax}
\tablenotetext{d}{Fulbright (2000)}
\tablenotetext{e}{Ryan, Norris \& Beers (2000)}
\tablenotetext{f}{Norris, Ryan, Beers \& Deliyannis (1997)}
\tablenotetext{g}{McWilliam, Preston, Sneden \& Searle (1995)}
\tablenotetext{h}{\teff\ based on analysis of Balmer line profiles.}
\tablenotetext{i}{\teff\ is deduced from the Fe spectrum.}
\tablenotetext{j}{Star assumed to be at/near main sequence turnoff.}
\label{table_oldabund}
\end{deluxetable}

%
%
\begin{deluxetable}{lrrrrr}
\tablenum{4}
\tablewidth{0pt}
\scriptsize
\tablecaption{Sensitivity of the Various Colors for EMP Turnoff Stars} 
\tablehead{
\colhead{Color} & \colhead{$1\sigma$\tablenotemark{a}} & 
\colhead{$\delta$(\teff)\tablenotemark{b}} & 
\colhead{$\delta$(\grav)\tablenotemark{b}} & 
\colhead{$\delta$([Fe/H])\tablenotemark{b}} & 
\colhead{$\delta[E(B-V)]$\tablenotemark{b}}   \nl
\colhead{} & \colhead{(mag)} & \colhead{(K)} & \colhead{(dex)} & \colhead{(dex)} &
\colhead{(mag)} \nl
}
\startdata
$B-V$ & 0.02 & $-125$ & +1.4 & $>2.0$ & +0.020 \nl
$V-J$ & 0.03 & $-71$ & +1.7 & $>2.0$ & +0.014 \nl
$V-K$ & 0.04 & $-73$ & +1.7 & $>2.0$ & +0.015 \nl
\enddata
\tablenotetext{a}{These are typical $1\sigma$ observational errors.}
\tablenotetext{b}{Change in the stellar parameter for a +1$\sigma$ change in the color.}
\label{table_sensitivity}
\end{deluxetable}

%
%
\begin{deluxetable}{l rr rrrl}
\tablenum{5}
\tablewidth{0pt}
\scriptsize
\tablecaption{Stellar Parameters for the Stellar Sample} 
\tablehead{
\colhead{Name} & \colhead{\teff($B-V)$} & \colhead{\teff($V-J)$} & 
\colhead{\teff($V-K)$} & 
\colhead{\teff(H$\delta$)} & \colhead{\teff(adopt)\tablenotemark{a}} & \colhead{\grav} \nl
\colhead{} & \colhead{(K)} & \colhead{(K)} & \colhead{(K)} &
\colhead{(K)} & \colhead{(dex)}  \nl
}
\startdata
Very Bright \nl
HD 140283 &  5935 & 5710 & 5785 &   5800 & 5750 & 3.67   \nl
BD+3~740 &  6750 & 6260 & 6450 &  6400 & 6355 & 4.0  \nl
G139--8 & 6500 & 6105 & 6290 &  6200 & 6200 & 4.5\tablenotemark{e}  \nl
  &  \nl
HK Stars \nl
BS 17447--029 & 6800 & 6470 & 6570 &  ... &  6530 & 4.4\tablenotemark{e}  \nl
CS 22878--101 & 4870 & 4735 & 4820 &  ...\tablenotemark{c} & 
        4775 & 2.3  \nl
CS 22950--046 & 4870 & 4695 & 4770 &  ...\tablenotemark{c} & 4730 & 2.3  \nl
  &  \nl
HES Stars \nl
HE~2133--1426 &  6475 & 6145 & 6460 &  6300 & 6300 & 4.1   \nl
HE~2344--2800\tablenotemark{b} & 6700 & 6500 & 6750  & 6400 & 6625 & 4.3  \nl
HE~0024--2523 & 6550 & 6700 & 6550  &  6500 & 6625 & 4.3  \nl
HE~0130--2303 & 6600 & 6600 & 6525  & 6500  & 6560 & 4.3  \nl
HE~0132--2439 & 5440 & 5250 & 5375 &  5400 & 5310 & 3.4   \nl
HE~0148--2611 & 6800 & 6500 & 6600 & 6500  & 6550 & 4.3  \nl
HE~0218--2738\tablenotemark{d} & 6625 & 6500 & 6600  & 6400
     & 6550 & 4.3  \nl
HE~0242--0732 & 6550 & 6360 & 6550\tablenotemark{f} &  6200 & 6455 & 4.2  \nl
\enddata
\tablenotetext{a}{The adopted \teff\ is a mean of that derived
from the dereddened $V-J$ and $V-K$ colors. See text.}
\tablenotetext{b}{This is a re-discovery of CS 22966--048, originally
found in the HK survey.}
\tablenotetext{c}{Star is too cool to show Balmer line wings.}
\tablenotetext{d}{The HIRES spectra show this star is a spectroscopic binary.}
\tablenotetext{e}{The lower luminosity solution is chosen, so the star is
fainter than the main sequence turnoff.  See text.}
\tablenotetext{f}{The uncertainty in the 2MASS $K$ mag is unusually 
large, $1\sigma$ = 0.10 mag.}
\label{table_param}
\end{deluxetable}

%
%
\begin{deluxetable}{l c rc rr}
\tablenum{6}
\tablewidth{0pt}
\scriptsize
\tablecaption{Comparison of $T_{eff}$ Determinations} 
\tablehead{
\colhead{Name} & \colhead{\teff}  & \colhead{$E(B-V)$} & \colhead{Ref.} 
& \colhead{\teff(Table 3)} & \colhead{\teff(Table 3 With}  \nl
\colhead{} & \colhead{(K)} &
\colhead{(mag)} & \colhead{}  & \colhead{(K)} & 
\colhead{Previous $E(B-V)$)\tablenotemark{a}} \nl
}
\startdata
Very Bright \nl
HD 140283 & 5680--5810 &  0.01 & b,c,d, e & 5750 & 5785 \nl
BD+3~740 & 6240 &  0.01 & f &  6355 & 6260 \nl
G139--8 & 6025 &  0.04 & g & 6200 & 6105 \nl
  &  \nl
HK Stars \nl
CS 22878--0101 & 4780 &  0.06 & h & 4775 & 4730 \nl
CS 22950--0046 & 4635 &  0.06 & h & 4730 & 4620 \nl
\enddata
\tablenotetext{a}{Here we derive \teff\ using our photometry 
and calibrations but adopt the reddening
value used in the previous abundance determination.}
\tablenotetext{b}{Ryan, Norris \& Beers (1996)}
\tablenotetext{c}{Nissen \etal\ (1994) and Edvardsson \etal\ (1994)}
\tablenotetext{d}{Korn \& Gehren (2000) -- using \grav\ from 
Hipparcos parallax}
\tablenotetext{e}{Fullbright (2000)}
\tablenotetext{f}{Ryan, Norris \& Beers (2000)}
\tablenotetext{g}{Norris, Ryan, Beers \& Deliyannis (1997)}
\tablenotetext{h}{McWilliam, Preston, Sneden \& Searle (1995)}
\label{table_tcompare}
\end{deluxetable}

%
%
\begin{deluxetable}{lrrr}
\tablenum{7}
\tablewidth{0pt}
\scriptsize
\tablecaption{Spectroscopic Parameters for the Stellar Sample} 
\tablehead{
\colhead{Name} & \colhead{$v_r$\tablenotemark{f}} & \colhead{[Fe/H](K)\tablenotemark{a}} & 
\colhead{[Fe/H] (HIRES)\tablenotemark{b}} \nl 
\colhead{} & \colhead{(\kms)} & \colhead{(dex)} & \colhead{(dex)}  \nl
}
\startdata
Very Bright \nl
HD 140283 &  $-171.0\tablenotemark{c}$  & ... & $-2.43$ \nl
BD+3~740 &  +174.5  & ... & $-2.69$ \nl
G139--8 & $-114.0$ & ... &  $-2.04$ \nl
  &  \nl
HK Stars \nl
BS 17447--029 & $-205.0$ & $-3.15(0.26)$ & $-2.91$ \nl
CS 22878--101 & $-131.7$ & $-2.91(0.18)$ & $-3.04$ \nl
CS 22950--046 & +106.7 & $-3.63(0.15)$ & $-3.28$ \nl
  &  \nl
HES Stars \nl
HE~2133--1426 &  +19.6 & $-3.32(0.23)$ & $-2.81$  \nl
HE~2344--2800 & $-135.8$ & $-3.06(0.20)$ & $-2.53$ \nl
HE~0024--2523\tablenotemark{d} & $-181.6$ & $-3.08(0.24)$ & $-2.62$ \nl
HE~0130--2303 & +74.2 & $-3.15(0.19)$ & $-2.93$ \nl
HE~0132--2439 & +294.7 & $-3.05(0.33)$ & $-3.56$  \nl
HE~0148--2611 & $-223.1$ & $-3.42(0.17)$ & $-2.96$ \nl
HE~0218--2738\tablenotemark{e} & +134.0 & $-3.81(0.17)$ & $-3.52$  \nl
HE~0242--0732 & $-188.9$ & $-3.59(0.20)$ & $-3.04$ \nl
\enddata
\tablenotetext{a}{These values are inferred from the 
strength of the Ca II K line and the Balmer lines in the 
moderate-resolution spectra.  There is a small systematic offset of $0.23$ dex
between the abundance scales of the followup spectra and of the high resolution
spectra.  See the text for details.}
\tablenotetext{b}{[Fe/H] values are from the detailed
abundance analysis presented in Paper II.}
\tablenotetext{c}{The uncertainty in the radial velocities is dominated
by systematic errors, and is $\pm1.0$ \kms.}
\tablenotetext{d}{This star has easily detectable CH absorption in the G band
and its lines are resolved.}
\tablenotetext{e}{The HIRES spectra show this star is a 
double-lined spectroscopic binary.}
\tablenotetext{f}{This is the heliocentric radial velocity.}
\label{table_specinfo}
\end{deluxetable}

\clearpage

%
%

\end{document}